\documentclass{acmart}
\AtBeginDocument{%
  \providecommand\BibTeX{{%
    \normalfont B\kern-0.5em{\scshape i\kern-0.25em b}\kern-0.8em\TeX}}}
\usepackage{xspace}
\usepackage{fancybox}
\usepackage{tcolorbox}
\tcbuselibrary{breakable}
\begin{document}

\newcommand{\TODO}[1]{{\color{red} TODO: #1}}
\newcommand{\chatbot}{LLM-based chatbot\xspace}
\newcommand{\chatbots}{LLM-based chatbots\xspace}

\newcommand{\custombox}[1]{
        \begin{center}
        \vspace{3pt}
        \Ovalbox{
        \begin{tcolorbox}
        \vspace{3pt}
             #1
        \end{tcolorbox}
        }
        \end{center}
}

\title{Revisiting Human Information Foraging: Adaptations for LLM-based Chatbots}

\author{Sruti Srinivasa Ragavan}
\email{srutis@cse.iitk.ac.in}
\affiliation{%
  \institution{Indian Institute of Technology}
  \city{Kanpur}
  \country{India}
}

\author{Mohammad Amin Alipour}
\affiliation{%
  \institution{University of Houston}
  \city{Houston, TX}
  \country{USA}
  }
\email{}


\begin{abstract}
  Information Foraging Theory's (IFT) framing of human information seeking choices as decision-theoretic cost-value judgments has successfully explained how people seek information among linked patches of information (e.g., linked webpages). However, the theory stands to be adopted and validated in non-patchy \chatbot environments, before its postulates can be reliably applied to the design of such chat-based information seeking environments. This paper is a thought experiment applying IFT's cost-value proposition to \chatbot environments and presents a set of preliminary hypotheses to guide future theory-building efforts for how people seek information in such environments. 
  
\end{abstract}

\maketitle

\section{Introduction}
The abundance and centrality of information in day-to-day tasks marks the information age we live in, and humans are dubbed "informavores" \cite{miller}. The new intelligent \chatbots (e.g. ChatGPT\footnote{https://chat.openai.com/}), have transformed how people search for information: they are able to comprehend users' query prompts for general-purpose information and use the interaction context to produce better quality responses. In turn, this has led to proliferation of new interactive systems in different domains (e.g., \cite{chen2023llm,zhou2023llm,liu2023internchat}). 
These systems seem to produce ``reasonable'' answers with arguably vague or incomplete prompt instructions, and seem to facilitate easy and quick acquisition and processing of information by their user\cite{cambonearly}.

However, it is unclear how information seekers adapt to these new interaction paradigms, such as cognitive adaptations to effectively learn and use these tools for information seeking \cite{anderson}, or to deal with their unique limitations (e.g., inaccuracies, nondeterminism) \cite{sarkar}, or the metacognitive effects (e.g., perception of acquired skills) are unclear \cite{eloundou2023gpts,matin-icer}. This lack of understanding warrants more fundamental inquiries into the nature of work with such systems--both by building new theories (e.g., \cite{barke}) and by revisiting relevant classical theories--for human information acquisition, information processing, and information refinement and retention. In this paper, we undertake one such exercise in the latter category and revisit a well-established theory in HCI, namely the Information Foraging Theory, in the context of these \chatbots.

Information Foraging Theory (IFT) is a theory of how people seek information. Proposed by Pirolli and Card in 1995,  IFT is an adaptation of the optimal food foraging theories in biology \cite{pirolli1995}. IFT posits that informavores forage for information in ways similar to how animals forage for their prey. Specifically, in both cases, the forager aims to solve an optimization problem: the foraging animal aims to optimize the rate of energy gain (i.e., energy gained per unit of energy expended hunting and consuming the prey), whereas the informavore attempts to maximize the rate of information gain (amount of valuable information gained per unit foraging cost).

The theory also describes how this cost-value optimization plays out in a specific kind of environment -- namely, where the information is organized as ``patches'', and the patches are interconnected by "links". In such environments, according to IFT, the forager follows a scent-following strategy, wherein they metaphorically "sniff on" (or attend to) "cues" that signal the value-cost payoffs of a foraging choice (e.g., text on hyperlink signals what is on the other end of the link). They then follow the path that smells strongest of the prey, much like the scent following behavior of animals. The theory also makes predictions about how much time a forager will spend on one information patch -- once again framing it in terms of the value obtained, as a function of cost (including the cost of lost opportunity in other costs).

Prior work has explored the validity of IFT in other types of information environments than on which the theory was developed (namely, interconnected filesystems and the web \cite{pirolli1995}). Examples include those with heterogeneous patches (text, images, animations, etc.) \cite{pirolli1995}, variationed environments having several variants (or versions) of the same information patches \cite{ragavan2016}, and in the case of screen readers \cite{vigo} such as for the visually impaired. Whereas in the former two cases, foragers followed scent albeit with some special adaptations by the forager, there were no scent following mechanisms with screen readers and perhaps even no cost-value optimization due to the design of the environments. Thus, with every new kind of information environment, there is a need to test the validity of the theory.

To this end, this work is a prelude to our ongoing research on validating IFT for \chatbot environments, especially those powered by Large Language Models (LLMs) (e.g., ChatGPT). Our key research questions are as follows:
\begin{itemize}
    \item Does IFT's cost-value optimization proposal apply to information seeking in\chatbots?
    \item If so:
    \begin{itemize}
    \item what specific cost-value optimization mechanisms do users adapt to effective forage for information in \chatbots?
    \item What do these mechanisms tell us about how to design \chatbot environments to better suit user needs?
    \end{itemize}
    \item If not:
    \begin{itemize}
    \item where does information seeking in \chatbots deviate from IFT, and what is another possile theory for such kinds of information seeking (or perhaps, more aggressively, is there a more unifying theory for human information seeking!). 
    \end{itemize}
\end{itemize}

In the rest of this paper, we revisit the roots and assumptions of Information Foraging Theory, discuss the nature of information seeking in QnA, in comparison with traditional notions of foraging (particularly in the "patchy" web), propose a list of initial hypotheses about human information foraging in \chatbots, and finally. We conclude with a discussion on the challenges and consequences of applying IFT to \chatbot contexts.
\section{Background: Information Foraging Theory?}

Information Foraging Theory (or IFT) concerns with how people seek information as part of a larger task. With roots in evolutionary psychology, IFT assumes that humans have exapted their food-foraging adaptations to aid their information seeking \cite{pirollibook}; thus IFT itself is an adaptation of the optimal foraging theory from biology \cite{stephenkrebs}. In fact, IFt uses the constructs \textit{predator} and \textit{prey} to refer to human information seekers or foragers, and to the information they are looking for, respectively. 

IFT has three key propositions (listed below). These propositions are not only descriptive--in that they explain \textit{how} humans forage for information, but also explanatory, in that they offer explanations for \textit{why} a forager might engage in the foraging behaviors they do. This explanatory power is then leverageable for practical purposes, such as to predict how a person would forage in a given environment (such as to evaluate its usability), or as a corollary to design information environments that are better suited to human foraging behaviors.

   \subsubsection{Value-cost optimization}: Central to IFT is the cost-value proposition which says that, analogous to predatory animals maximizing the rate of energy gain during foraging, \textit{informavores} foraging for information adopt mechanisms to optimize the rate of information gain (or the amount of valuable information per unit cost). It is important to note that despite efforts, foragers might not always be successful in their cost-value optimization attempts, and this is largely due to incomplete information about their environment (for example, the cheapest way to get to an information piece, or the ability to estimate what information is in a webpage). In other words, there is a distinction between predator's estimates or expectations of costs and values from various foraging choices, and the actual costs and values in the foraging; only when the estimates and actual match do a forager actually engage in optimal user experiences and efficient foraging.

    \subsubsection{Patch models}: The second set of IFT propositions assumes that information in the environment is organized as ``patches'' (e.g., web pages, files). In turn, these patches could form a hierarchy (e.g., Internet contains several domains containing many subdomains containing many websites containing many pages each containing sections, etc.). In fact, such hierarchies are a common way in which complexity is organized in several information environments (folder structure in computers, sections in large documents, websites, sections, and chapters in books and libraries). 
    
    By laws of evolutionary psychology, humans adapt themselves to their environment, and in this case, they have adaptation mechanisms to optimally forage for information in such patchy heirarchies. The patch models of IFT seek to explain what these adaptations are.
    
    \begin{itemize}
        \item \textit{Between-patch foraging times}: In a patchy information environment, a predator must choose among several available patches (e.g., which page among hundreds of search results), so as to maximize the rate of information gain. For this \textit{between-patch foraging}, IFT says, a predator relies on the \textit{scent} from various \textit{cues} in the environment, and follows the path with strongest scent of prey. For example, in web search results, a person might look at the ratings of a webpage, or look at its preview or a snippet to guess what the page contains, and whether to click on the link or not. 
        
        \item \textit{Within-patch foraging times}: Having navigated to a patch, the forager must then gather information foraging within that patch (including any sub-patches therein). How long a predator spends doing so, before leaving the page for another one, is dependent on the average rate of gain of information. The longer a predator spends in a page, the more information s/he would consume, thereby diminishing the value of a page. Thus, a predator will remain in a patch until the average rate of information gain in the current patch is greater than what s/he expects to get from find, navigating to and foraging in another patch.
\end{itemize}
    
    Indeed, both these activities are tedious when a predator must seek very specific information in too many large patches. Therefore, another cost-value optimization strategy is called \textit{"enrichment"}. Here, a predator aims to improve the value or lower the cost of foraging by engaging in activities such as search, filtering, sorting, or even customizing the environment (e.g., split tab to constantly lookup something, filter to get only relevant information). 
   
\subsubsection{Diet model}: The third proposition of IFT is the diet model. This model also describes cost-value adaptations of information foragers, but cases where there are multiple possible prey(s) or prey kind(s) available (for example, variety of news, emails, books to choose from). In such cases, IFT predicts what kinds of "prey" a predator chooses to be part of the optimal diet for the forager: that a \textit{rational} forager will select high-value prey(s) and leave out low-value prey(s), irrespective of cost, assuming similar occurrence of both high- and low-value prey(s). For example, a rational and task-oriented person would never read what he/she knows to be a spam email (unless the task itself has to do with reading the spam email). In fact, there has even been recent work on what diet is part of the foraging of AI users when seeking explanations about the output of an AI system, as part of efforts in eXplainable AI (XAI) systems \cite{dodge}. 

\section{Information Foraging: Web vs. \chatbots}

\subsection{Information seeking on the Web}
We mentioned earlier that the World Wide Web (WWW) is a patchy structure, and that the patches for a hierarchy: for example, web domains contain subdomains contain websites contain web pages contain sections. 
response,that Webby, a new college graduate, is foraging the WWW to learn about mutual funds and investments after earning her first paycheck. She starts by going to an Internet Search Engine (e.g., Google \footnote{www.google.com}) and types in "mutual funds". She receives a set of links to various pages that contain information relevant to her search query. Webby then engages in between-patch foraging via scent-following: she skims through the search results, paying attention to the link title, the snippets of content below them, or the reputation of the site and her past experiences with it, and clicks on one that seems potentially useful (high value) and/or easy to grasp (low cost).

Once on a page, Webby then reads the content skimming, scanning, and skipping ahead as needed--all ways of reducing cost. Then, when she feels she has gathered enough from the page, she leaves to go back to the search results page, and then the cycle continues. Sometimes, while on a page, Webby finds a link that seems promising, and she clicks on it, and other times, she lands on a page and is disappointed at its content and immediately abandons the patch.

At this point, it is easy to see that all these behaviors are cost-value optimizations--the mechanisms of which are explained by IFT's patch and diet models \cite{pirolli1995}.

\subsection{Information seeking with \chatbots}
Now, let us consider the case of another forager, Chad. Like Webby, Chad also wants to start investing in mutual funds and first wants to learn about them. But unlike Webby, Chad goes to ChatGPT and prompts it with a question: ``What are mutual funds?''. In response, the \chatbot provides a detailed answer about what mutual funds are. Both Chad and the \chatbot go back and forth until he feels he has learned about mutual funds.

Chad's interactions with ChatGPT look as follows: 
\begin{itemize}
    \item Sometimes, Chad gets lucky and gets the right answer the very first time, and is able to grok the response quickly and move on. 
    \item Often, Chad reads the \chatbot's response to a prompt and decides that it is not what he wanted, and therefore reframes his prompt to better specify his needs.
    \item He sometimes also finds the content too hard to grasp (costly) or not quite relevant, and so he refines the prompt, or asks a follow-up mimicking human-human conversation (e.g., "can you put it in simpler terms?", "explain", "what are mutual funds in simple terms?", "oh, I meant apple the computer and not the company").
    \item Occasionally, Chad might see something suspicious or hard to believe, and might ask the chatbot for further clarifications, reasoning, or evidence (or check its veracity elsewhere).
    \item Sometimes, Nash also wants to go back to a thread he left, but repeating the same prompt or something similar does not always guarantee the same answer, and so he cannot exactly pick up from where he left. He tries and gives up, and plows through the chat history to go back to where he left. 
\end{itemize}

All of the above behaviors described for Chad are based on our ongoing lab study comparing information seeking between Web and \chatbot, and from prior studies (e.g., \cite{skjuve2023, menon2023}). 

\subsection{Information seeking similarities: Web vs. \chatbots}
Comparing information seeking on the Web and in \chatbots, several similarities and differences become apparent; the similarities are described in this section, and the differences are described in the next section. 

\subsubsection{Similarity \#1: Forager specifies foraging goals.}
First, in both the Web and the Q\&A chatbot, the information seeking starts with a forager specifying the foraging goal. Whereas in the search engine, this is typically keywords in the foraging goal, the expected prompt in the chatbot is a question, such as in human-human conversations. \footnote{Indeed, sometimes a web forager does not use a search and instead directly goes to a website and follows the scent to their prey--however, that is arguably less common that using a search engine.}

\subsubsection{Similarity \#2: Foraging goals evolve.}
Second, in the case of both the web and \chatbot, the foraging goal evolves. This could be in the form of seeking clarifications about some information the forager had just gathered, or asking follow-up questions that in some cases can be formed as the next subtask, based on what has been found. Both environments cater to such kinds of foraging -- by way of refining the query, or asking a follow-up question (e.g., ``what does X mean?'', ``why do you say X?''), or sometimes by clicking on a link on the Web (such as hovering over a word for meaning, or citation). In fact, one might argue that this is a fundamental characteristic of information seeking, when the prey is ill-defined. 

\subsubsection{Similarity \#3: Value-cost judgments guide foraging.}
Third, information seeking in both the Web and \chatbot is driven by value-cost judgments. For example, on the web, participants seek pages that seem easy to understand based on snippets of search results. Similarly, in the chatbot environment, participants read the first few lines of the response, or eyeball the structure and length, to estimate whether the response is relevant, or too hard, and then seek ways to optimize it (for example, reframing prompt to simplify something). 

\custombox{
\textbf{Similarities: Web vs. \chatbot foraging.}
\begin{itemize}
    \item Foragers begin foraging by specifying a verbal description of the prey. 
    \item Foraging goals evolve as the forager gathers information, as well as in response to the environment. 
    \item The fundamental cost-value optimization (i.e., maximizing value-to-cost ratio) is central to what drives foraging behaviors in both environments.
\end{itemize}
}
\subsection{Information seeking differences: Web vs. \chatbots}
In terms of differences, we observe distinctions in the environment as well as foraging behaviors, and we frame them as hypotheses (within the framework of IFT) that we aim to empirically verify, dismiss or refine, in future work. 

\subsubsection{Difference \#1: Prey specification.} 
Both in the preliminary results of our study, as well as in prior work \cite{sarkar2022}, there is some evidence that new users of \chatbots initially treat them similar to web search engines, and provide search keywords --  eventually gravitating towards asking questions that the \chatbot environments are built for. In other words, in an \chatbot environment, a user is \emph{expected to know} what they are looking for and to specify it clearly (e.g., "where is Paris located?"), rather than just words about which they might be interested (e.g., "Paris" or "Paris location"). Framed in terms of costs and values, it therefore appears that the former questioning of chatbots (e.g., via prompt engineering) is harder than keyword-based web search. 

\custombox{\textit{Hypothesis 1:} The cost of specifying prey, or enrichment to improve overall value of information in the environment, is higher in the  \chatbot environment (via prompt engineering) than in the web environment (via web searches).}


\subsubsection{Difference \#2: Structure of the information environment.} 
A second point of difference between the web and the chat environment is based on the structure. We already saw that the web is organized as hierarchical patches, and that the prey lives in one or more of those patches. In contrast, the chat is one long patch, that evolves and changes with each interaction between the user and the \chatbot environment. 
Notably, as each new question is asked and responses are generated, the new information is appended to the same patch. One could argue that each question and answer are a smaller patch, but there are no explicit boundaries (e.g., each question-response pair can be considered a patch, but what about follow-up questions?). 
Moreover, there are also limited enrichment options (e.g., filtering out irrelevant patches, or removing them) compared to traditional patchy environments. In fact, even considering other human-human conversations, the resultant transcript is perhaps considered as one single patch, rather as multiple pages on the web, or files in a folder. However, whether users perceive chat as one large "blob" or as logical patches remains to be empirically understood. 

\custombox{\textit{Hypothesis 2:} Whereas users treat the \chatbot environment has a single \textit{constantly evolving} patch, the web environment is a collection of multiple patches of varying size}

\subsubsection{Difference \#3: Transience of patches.} In WWW, the prey is nearly "permanent"; a forager can find their way to the same prey, any number of time times, such as by bookmarking, or leaving an open tab, or a new similar web search -- both within the same session and another future session. In contrast, in ChatGPT or other similar \chatbot environments, the prey is \textit{ephemeral} -- it is created on the fly in response to a user prompt and context, a user might not be able to get the same prey by typing in the same prompt. They might have to go through the tedium of finding it in chat history (provided that it is saved, and not too long!).

This has notable consequences for cost-value estimation. First, some prey might be very hard or impossible to get; this can be framed as infinite cost, or zero availability of the prey. In driving this down, a predator might still engage in very costly forms of foraging, such as scrolling through chat history, constantly remembering to bookmark, or spending effort in prompt engineering or finally adapting and reorienting themselves to a new prompt's response. Second, repeated experiences of this nature might affect the users' estimations of costs and values in the future, and as we shall see later in this section, trust has a role to play in this regard. 

\custombox{\textit{Hypothesis 3:} When prey is scattered across responses or patches, the overall foraging costs are higher in \chatbot environments than in the web.}. 

\subsubsection{Difference \#4: What information is available.}
Another distinction is in what is in the information environment. In the web, there exist several patches that a predator must forage to piece together the prey. There might even be multiple alternative prey and prey types (e.g., video, text, audio) and a predator has access to all of this prey, and can even see the presence of what else might be available (even when they are not foraging in a patch). In other words, a predator is aware of, and makes decisions about what opportunities to pursue, and what opportunities they lose and at what cost. 

In contrast, in the chat environment, there is a specification of the prey, and then one instance of the prey presented in one coherent manner. This has two consequences: 

\begin{itemize}
    \item The forager does not know what other alternative prey(s) exist, and therefore a predator has no sense of opportunity cost, nor knows whether what they actually get as response is really the most profitable.

    \item The lack of diversity in prey results in a lack of diversity in diet (e.g., multiple viewpoints). 
\end{itemize}

\custombox{\textit{Hypothesis 4:} Foraging in \chatbots blinds foragers  to other, more profitable and diverse diets compared to the web.}

\subsubsection{Difference \#5: Cost-value judgment adaptations}
This also brings us to the next difference in foraging. When foraging in the web, a forager has different patches to pick from; they consider the costs and values of those patches before making a decision to navigate into one and consume its content or not. This is done by scent-following: where scent is a predator's estimate of what is in a patch even before getting there. 

Then, once they enter a patch, they do another evaluation (such as by seeing length of page, or nature of content) to see whether the actual costs and values are as per their prior expectations, and if not, they leave the patch, and readjust their expectations of costs and values of that patch for future foraging decisions.

In contrast, in \chatbot environments, there are no diverse patches to choose from, and a predator has no way of gathering diverse prey(s) or even knowing that diversity in prey(s) exists. Moreover, there is not even a real patch whose cost and value a forager can estimate before deciding whether to navigate or not. We believe that the user's adaptation to such ecological uncertainty is \textit{trust}. 

Before performing the action of what to prompt the chatbot with, and what to expect, the predator has a certain level of trust in the \chatbot. We hypothesize that this trust is effectively a predator's expectation of cost and value (e.g., based on hearsay, reputation, past experiences) about what to expect from the \chatbot, or what prompt would generate what response, and with what accuracy. Then, once they do generate the patch, they then evaluate the content for both relevance and accuracy -- based on the cost-value payoffs expected (cost of errors, prior trust)-- in turn recalibrating their expectations of costs and values for future foraging.

\custombox{
\textit{Hypotheses 5: 
} \begin{itemize}
     \item  Cost-value  estimations in \chatbots include estimations of relevance and accuracy. 
     \item To cope with high levels of uncertainty and lack of other patches in \chatbot environments, foragers use \textit{trust} to make preliminary cost-value estimates, such as for what response a prompt will provide, or how much to trust a response. (This is analogous to scent following adaptation for too many patches in the web!). 
     
     \item Once within a patch, a forager will evaluate costs and values, both in terms of relevance, as well as accuracy (just as they might do on the web); this investment is based on the potential costs of mistakes. This happens by skimming through the content, presence of cues such as such as citations, seeking additional information (e.g., counterfactuals and explanations in chatbot or fact checking in another environment). 
     \item Foragers' trust (cost-value estimates based on past experiences) in the \chatbot acts as a moderator (and not a mediator) in how much cost the forager is willing to expend for accuracy estimation. 
    \item  Trust (similar to scent in the web) is foragers' ability to estimate the value and cost to be gained from that environment (e.g., based on knowledge and prior experience). Overtrust and undertrust, are simply misjudgments of such costs and values.  
   
\end{itemize}
}

\section{Future directions}
Indeed, the above hypotheses are preliminary and un-verified. Our intention then with these verbal hypotheses is as a first step in theory building. Our future steps plans include: mathematical formulation of verifiable hypotheses, combining them to form propositions, evaluating their explanatory and predictive abilities empirically, and finally building a theory that fits easily into the existing information-foraging theoretic framework, or as a standalone theory for information seeking with \chatbots. 

\section{Concluding Remarks}
In this paper, we reported our ongoing efforts to adapt the information foraging theory for \chatbots. We compared classic foraging in the web and in \chatbots and discussed how current theory may be inadequate to be applied to this new domain.
We proposed five sets of hypotheses to augment or modify the current theory. In particular, we discussed novel characteristics of patches and scent-following behaviors in such systems, and the role of trust as an ecological adaptation of human foragers in \chatbot environments. We are conducting studies to rigorously validate these hypotheses.  

\bibliographystyle{ACM-reference-format}
\bibliography{refs}

\end{document}